%% file: main.tex
\documentclass[sigconf]{acmart}

\usepackage{caption}
\usepackage{subcaption}

\AtBeginDocument{%
  \providecommand\BibTeX{{%
    \normalfont B\kern-0.5em{\scshape i\kern-0.25em b}\kern-0.8em\TeX}}}

\setcopyright{acmcopyright}
\copyrightyear{2018}
\acmYear{2018}
\acmDOI{10.1145/1122445.1122456}

\acmConference[EVIA '23]{EVIA '23: The 10th International Workshop on Evaluating Information Access}{December 12-15, 2023}{Tokyo, Japan}
\acmBooktitle{Woodstock '18: ACM Symposium on Neural Gaze Detection,
  June 03--05, 2018, Woodstock, NY}
\acmPrice{15.00}
\acmISBN{978-1-4503-XXXX-X/18/06}

\renewcommand\footnotetextcopyrightpermission[1]{} 
\begin{document}

\title{Decoy Effect in Search Interaction: A Pilot Study}

\author{Nuo Chen}
\email{pleviumtan@toki.waseda.jp}
\affiliation{%
  \institution{Waseda University}
  \city{Tokyo}
  \country{Japan}}

\author{Jiqun Liu}
\email{jiqunliu@ou.edu}
\affiliation{%
  \institution{The University of Oklahoma}
  \city{Norman}
  \country{OK, USA}}
  
\author{Tetsuya Sakai}
\email{tetsuyasakai@acm.org}
\affiliation{%
  \institution{Waseda University}
  \city{Tokyo}
  \country{Japan}}

\author{Xiao-Ming Wu}
\email{xiao-ming.wu@polyu.edu.hk}
\affiliation{%
  \institution{The Hong Kong Polytechnic University}
  \city{Hong Kong}
  \country{China}
}

\begin{abstract}
In recent years, the influence of cognitive effects and biases on users' thinking, behaving, and decision-making has garnered increasing attention in the field of interactive information retrieval. The decoy effect, one of the main empirically confirmed cognitive biases, refers to the shift in preference between two choices when a third option (the decoy) which is inferior to one of the initial choices is introduced. However, it is not clear how the decoy effect influences user interactions with and evaluations on Search Engine Result Pages (SERPs). To bridge this gap, our study  seeks to understand how the decoy effect at the document level influences users' interaction behaviors on SERPs, such as clicks, dwell time, and usefulness perceptions. We conducted experiments on two publicly available user behavior datasets and the findings reveal that, compared to cases where no decoy is present, the probability of a document being clicked could be improved and its usefulness score could be higher, should there be a decoy associated with the document.
\end{abstract}



\maketitle
\input{intro}

\input{relatedwork}
\input{rq}
\input{experiment}

\input{conclusion}

\section*{Acknowledgement}
We thank Prof. Kana Shimizu for providing valuable feedback for this study. We also thank all reviewers of this paper for their valuable feedback and constructive criticism. Jiqun Liu's participation in this project is partially supported by the Bridge Fuding Investment Program (BFIP) Award from the University of Oklahoma Office of the Vice President for Research and Partnerships. 
\balance
\bibliographystyle{ACM-Reference-Format}
\bibliography{decoy-ref}
\end{document}

%% file: intro.tex
\section{Introduction}

Understanding how users think, behave, and make decisions during interactions with search systems represents a foundational research interest in the field of interactive Information Retrieval (IR). A cognitive bias is a systematic pattern of deviations in thinking which may lead to irrational judgements and problematic decision-making~\cite{Tversky1974, Tversky1992}. In recent years, the exploration of cognitive biases and their impact on the information seeking and retrieval behaviors and outcomes has garnered increasing attention~\cite{liu2023behavioral, liu2019investigating, Azzopardi2021}. 

The \textit{decoy effect}, which is one kind of cognitive bias, describes a situation in which individuals alter their preference between two initial choices when introduced to a third~(\textit{i.e.} the decoy), which is asymmetrically inferior to one of the initial choices~\citep{Huber1982}. Figure~\ref{fig:water} illustrates an example of the decoy effect in shopping decision-making. In a shop, a customer who wants to buy a drink might waver between a 500ml bottle of water~(for $\$1.19$) and a bottle of soda with a similar size (for $\$1.49$). The 500ml water is cheap, but the soda tastes better, so it might be hard to make a decision and which one the cosumer will choose might depend on the assessment of their relative utility. Yet, with a  250ml  bottle of water for $\$1.09$ presenting beside the 500ml water, the customer might lean towards the 500ml water, as they perceive a substantial relative gain from the comparison of the 500ml water and the 250ml water:~spending an additional $\$0.10$ to purchase a  500ml bottle of water, compared to the 250ml one, evidently presents a highly economical deal. In the above example, the 250ml water serves as the \textit{decoy} to the \textit{target} 500ml water.

\begin{figure}[t]
    \centering
    \label{fig:water}
    \includegraphics[width=.7\linewidth]{
    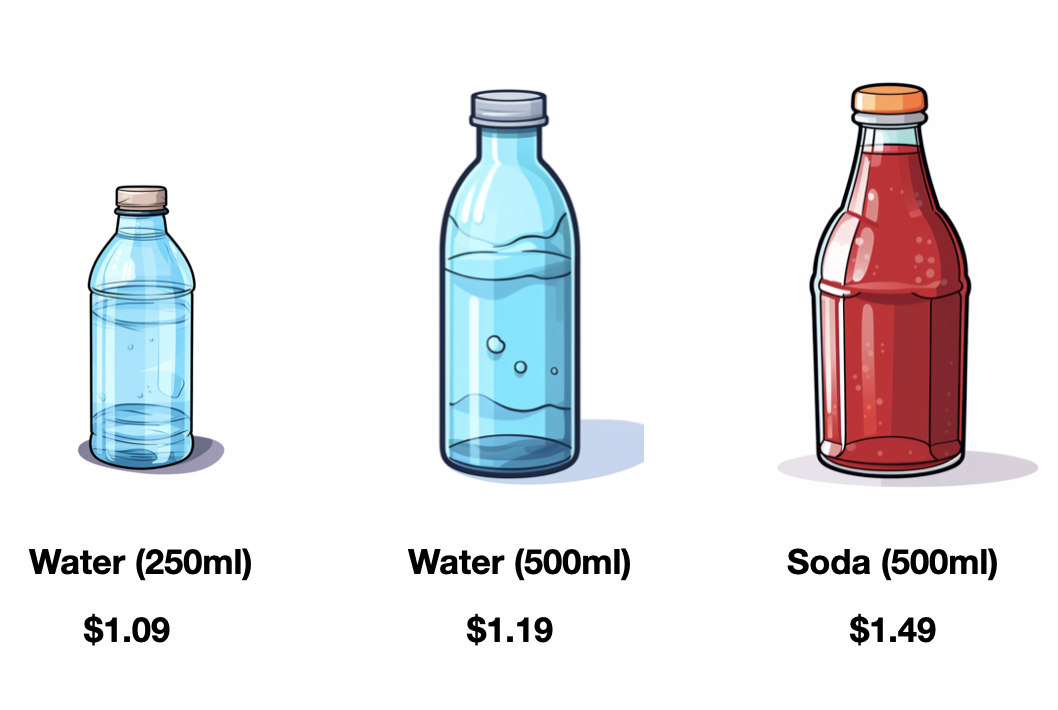
    }
    \caption{An example of the decoy effect. The image is generated by the authors using Midjourney.}
    \label{fig:water}
\end{figure}

In the field of information retrieval, Eickhoff~\cite{Eickhoff2018} examined the impact of a decoy document on  thresholds and strategies in crowdsourcing relevance judgments, showing that assessors could increase the relevance rating of target document when it is shown with the decoy document. Nevertheless,~\citet{Eickhoff2018} only focuses on crowdsourcing assessors operating within the annotation interface and few research currently addresses how the decoy effect influences user interactions on Search Engine Result Pages (SERPs).

 To address this gap, our study seeks to understand how the decoy effect at the document level influences users' interaction behaviors on SERPs, such as clicks, browsing dwell time, and usefulness perceptions. We conducted experiments on two publicly available user behavior datasets and the experimental results reveal that, compared to cases where no decoy is present, the probability of a document being clicked could be improved and its usefulness score could be higher, should there be a decoy associated with the document.

This study, through the perspective of behavioral economics, delves deeper into the behavioral patterns and decision-making processes of users interacting with search engines. The findings stand to encourage IR researchers to increase the explanatory power of formal search models from a realistic behavioral and psychological foundation.

%% file: relatedwork.tex
\section{Related Work}
Insights from cognitive psychology and behavioral economics suggest that, \textit{cognitive biases} arise from one’s limited cognitive ability when there are no enough resources to properly collect and process available information~\cite{kruglanski1983}. Due to cognitive biases, one’s decisions under uncertainty can systematically deviate from what is expected given rational decision-making models~\cite{Tversky1974, Tversky1991, Tversky1992}.

In search and recommendation contexts, interactions between individuals and systems could lead to the incorporation of behavioral signals, influenced by \textit{cognitive biases}, into datasets used for training machine learning algorithms, thereby potentially magnifying existing system biases~\cite{Azzopardi2021, liu2023behavioral}. Cognitive biases might also result in significant deviations in users' behaviors and judgements from optimal or desired outcomes. Consequently, this could give rise to unfair decisions and outcomes between users who are more susceptible to certain biases and contextual triggers and those who are not~\cite{liu2023}. Therefore,  with an increasing number of individuals turning to search and recommendation systems to access and utilize information for life decisions, the influence of cognitive biases on the information retrieval process is drawing heightened attention from IR researchers. Thus far, a lot of studies have explored the influence of cognitive biases such as the anchoring effect~\cite{Shokouhi2015}, the priming effect~\cite{Scholer2013, chandar2012}, the ordering effect~\cite{BANSBACK2014}, and the reference dependence effect~\cite{Liu2020, wang2023investigating} on document examination, relevance judgment, and evaluation of whole-session search satisfaction. In recent years, with the growing knowledge about users’ cognitive biases, some works began to introduce cognitive biases into the construction and meta-evaluation of evaluation metrics~\cite{zhang2020,chen2022,chen2023, liu2022matching}, but this is out of the scope of this paper.

In this paper, we specifically shed light on one of the cognitive effects, the \textit{decoy effect}. The \textit{decoy effect}, which is one kind of cognitive biases, describes a situation in which individuals alter their preference between two initial choices when introduced to a third~(\textit{i.e.} the decoy), which is asymmetrically inferior to one of the initial choice~\cite{Huber1982}. In the fields of e-commerce and recommendation systems, there have been some studies exploring the impact of the decoy effect~\cite{Teppan2010,teppan2015, mo2022decoy}. In the field of information retrieval, it is not clear how the decoy effect influences user interactions with and evaluations on Search
Engine Result Pages. The work most closely related to ours in theme is that of Eickhoff~\cite{Eickhoff2018}, which shows that when a relevant item is presented alongside two non-relevant items, with one non-relevant item being distinctly inferior (i.e., the decoy), assessors tend to rate the superior non-relevant document as more relevant. Nevertheless,~\citet{Eickhoff2018} only focuses on crowdsourcing assessors operating within the annotation interface and our study addresses how the decoy effect could influence user interactions on Search Engine Result Pages (SERPs), which is a broader scenario. 

%% file: rq.tex
\section{Research Question}

To address the research gap mentioned above, in this study, we seek to understand how the decoy
effect at the document level influences users’ interaction behaviors on SERPs, such as clicks, browsing dwell time, and usefulness perceptions. Specifically, our work sought to answer following \textbf{research question} (RQ):

\textbf{How, and to what extent, the presence of a decoy influences the likelihood of a document
being clicked, the browsing duration on it, and its perceived usefulness?}

%% file: experiment.tex
\input{fig_data_process}
\input{fig_feat}

\input{fig_dist}

\section{Experiment}
In this section, we introduce the datasets employed and the methodology adopted in the experiment, as well as the experimental results. 
\subsection{Datasets}
The THUIR2016 dataset~\cite{mao2016} is collected under controlled laboratory setting, whereby participants were instructed to execute a given intricate search tasks utilizing commercial search engines. This dataset encompasses a total of $9$ topics, $225$ search sessions and $933$ queries, along with the title and snippets on the SERP of each query. It also contains $4$-level user self-rating usefulness scores for the items they clicked and $5$-level graded relevance labels collected from external assessors. 

The THU-KDD dataset~\cite{Liu2019} is also collected under controlled laboratory setting similar to the THUIR2016 dataset. This dataset encompasses a total of $9$ topics, $450$ search sessions and around $1100$ queries, along with the title and snippets on the SERP of each query. It also contains $4$-level user self-rating usefulness scores for the items they clicked and $4$-level graded relevance labels collected from external assessors. 

\subsection{Data Processing Approach}
To find out potential decoy instances from user logs, we first proffer a definition of a decoy instance. A pair of documents, composed of the \textit{target document} and the \textit{decoy document} $(t, d)$, constitutes a decoy instance if and only if the following conditions are met: (1) $t$ and $d$ share certain degree of similarity in content, albeit not identical, i.e., $S_{\min} \leq \text{similarity}(t, d) \leq S_{\max}$, where $S_{\min}$ and $S_{\max}$ respectively represent the minimum and maximum similarity thresholds; (2) $d$ is inferior in quality to $t$, i.e., $\text{quality}(t) > \text{quality}(d)$; (3) the position $t$ and $d$ within a SERP is close enough, i. e., $|rank(t)-rank(d)| \leq \Delta_{\mathrm{rank}}$. 

In our experiments, given that the documents are all in Chinese, we first concatenate the title and snippet into a string. Subsequently, we employ a tool named \texttt{jieba}\footnote{https://github.com/fxsjy/jieba} to carry out word segmentation, obtaining a token list. We then calculated the cosine similarity with the vanilla definition~\cite{HAN201239} between each pair of documents under the same topic. We designated $S_{\min}$ as the 99th percentile of document similarity in each dataset, setting $S_{\max}$ to $0.95$. In THUIR2016 dataset, $S_{\min}$ stands at $0.626$ and in THU-KDD dataset, $S_{\min}$ stands at $0.594$. For the second condition, we employ the relevance scores given by external assessors as the measurement of document quality, mandating that $\text{relevance}(t) - \text{relevance}(d) \geq 2$ to ensure that the decoy is substantially inferior to the target. For the third condition, we require the absolute value of the difference of the rank between $t$ and $d$ is smaller than or equal to 5 ($\Delta_{\mathrm{rank}} = 5$). We processed the top 10 documents in each SERP on all datasets adhering to the aforementioned three conditions, and we identified $982$ records of decoy pairs involving $318$ distinct target documents in the THUIR2016 dataset; and $922$ records of decoy pairs involving $376$ distinct target documents. in the THU-KDD dataset. In the following discourse, we denote the set consisting of all target documents in a corpus as $\mathcal{T}$.

To investigate whether user interactions with documents are disparate when no decoy is present compared to situations with a decoy, we assign some documents not in  $\mathcal{T}$ to the control group~(\textit{i.e.,} \textit{control documents}), adhering to the following condition: A document $c$ which is not in the set of target documents~(\textit{i.e.,} $c \notin \mathcal{T}$) is considered a control document if and only if it matches a target document $t \in \mathcal{T}$ such that ~$\text{similarity}(c,t) \geq S_{\mathbf{control}}$ and ~$|\text{relevance}(c) - \text{relevance}(t)| <= 2$. We denote the set of all such $c$ as $\mathcal{C}$, and the set of all $t$ that can match with at least one $c$ as $\mathcal{T}^\prime$, where $ \mathcal{T}^\prime \subset \mathcal{T}$. In our experiments, we set $S_{\mathbf{control}}$ to the $99.5$th percentile of document similarity in each dataset. In THUIR2016 dataset, $S_{\mathbf{control}}$ stands at $0.709$ and in THU-KDD dataset, $S_{\mathbf{control}}$ stands at $0.676$. According to the aforementioned condition, we have identified $741$ qualifying \textit{control documents} in the THUIR2016 dataset and $1790$ in the THU-KDD dataset. 

We then extracted interaction records of all \textit{control documents} in the THUIR2016 and THU-KDD datasets, obtaining $1384$ and $2770$ records respectively. Subsequently, from the records of decoy pairs in the THUIR2016 and THU-KDD datasets~(with $982$ and $922$ records respectively), we filter out all records where $t \in \mathcal{T}^\prime$, obtaining $739$ and $828$ records respectively. Note that, for decoy pairs from the same SERP interaction $i$, there could be situations where the same target document corresponds to multiple decoy documents. In our filtering process, we ensure that for a given SERP interaction $i$ and a given target document $t$, only one record is eventually extracted. We concatenate the interaction records of target documents and control documents, ultimately obtaining document interaction record lists of lengths $2123$ and $3598$ in the two respective datasets. These two lists of interactions will be employed for the subsequent data analysis. In the subsequent analysis, we process the interaction signals as follows: for documents that have not been clicked, their usefulness score is assigned a value of 0, and their browsing duration is also set to 0. Figure~\ref{fig:data_process} provides a brief outline of our data processing workflow.

\subsection{Data Analysis and Experimental Results}

From Figure~\ref{fig:feat}, one can see that in both datasets, there are more instances receiving a score of $4$ (the maximum) in \textit{usefulness}. On the THU-KDD dataset, the target group exhibited fewer instances labeled as 0 and 1; the proportions of instances labeled as 2 and 3 were approximately similar between the target and control groups. On the THUIR2016 dataset, the target group exhibited fewer instances labeled as 1 and 2. Additionally, the proportions of instances labeled as 0 and 3 were approximately similar between the target and control groups. For \textit{click probability}, there is no difference in the THUIR2016 dataset, while in the THU-KDD dataset, the click probability of target group is higher. Regarding \textit{browsing duration}, on the THUIR2016 dataset, the target group exhibited a higher proportion of instances with browsing times ranging from 30 to 60 seconds. Apart from this observation, the distributions of browsing durations for both the target and control groups were approximately similar. On the THU-KDD dataset, the target group exhibited a higher proportion of instances with browsing times ranging between 0 to 5 seconds and 30 to 60 seconds, while a smaller proportion of instances had browsing times between 5 to 30 seconds.

From Figure~\ref{fig:feat}, it is challenging to derive an intuitive conclusion, especially regarding the relationship between the decoy effect and variables such as click probability, browsing duration and usefulness. Nevertheless, Figure~\ref{fig:dist} shows that, across both datasets, the distribution of target documents and control documents over ranks diverges. Hence, it is necessary to factor out any latent effects stemming from position biases on our results in order to draw a conclusion on the relationship between the presence of decoy and users' behavior. 

To control the impact caused by \textit{rank position}, we employ regression analysis to investigate the relationships between the presence of a decoy and the probability of clicks, browsing duration, and usefulness scores. We constructed three regression models, taking whether the document is clicked (\texttt{is\_clicked}), browsing duration  (\texttt{duration}), and usefulness score (\texttt{usefulness}) as dependent variables respectively, and the presence of a decoy (\texttt{has\_decoy}), the document's rank (\texttt{rank}), task ID (\texttt{task\_id}), and user ID (\texttt{user\_id}) as independent variables.

Note that contrary to computer science, in econometrics, regression models are predominantly employed for interpretation rather than for prediction. In a multiple regression model, each coefficient tells people the impact on the dependent variable of a one-unit change in that independent variable, holding all other independent variables constant~\citep{SMITH2012}. In this study, we foucus on elucidating how, and to what extent, the presence of a \textit{decoy} influences the likelihood of a document being clicked, the browsing duration on it, and its perceived usefulness. Hence, we do not partition the dataset into training and test subsets; instead, we perform regression on the entirety of the data. Including \texttt{rank}, \texttt{task\_id}, and \texttt{user\_id} as independent variables serves to use them as control variables to mitigate the potential influences from rank position, task type, and individual characteristics on the outcomes, thus better elucidating how variations in \texttt{has\_decoy} would affect the values of \texttt{is\_clicked}, \texttt{duration}, and \texttt{usefulness}. For \texttt{is\_clicked}, we employ Logistic regression, and for \texttt{duration} and \texttt{usefulness}, we resort to Ordinary Least Squares (OLS) regression. 

Table~\ref{table:reg} shows the regression coefficient of the independent variable \texttt{has\_decoy} with the dependent variables \texttt{is\_clicked}, \texttt{duration} and \texttt{usefulness}. As previously stated,   our focus in this research is to elucidate in what manner and to what extent the presence of a decoy impacts whether a document is clicked, the browsing duration, and the usefulness scores, and \texttt{rank}, \texttt{task\_id}, and \texttt{user\_id} are included merely to control for the effects brought by rank position, task, and individual characteristic respectively. Therefore, we opt to omit the reporting of the constant as well as the regression coefficients of \texttt{rank}, \texttt{task\_id}, and \texttt{user\_id} in the tables.

From Table~\ref{table:reg}, one can observe that: across the two datasets, the presence of a \textit{decoy} could exert a positive influence on the likelihood of being clicked (coefficient = $0.363$ and $0.217$ respectively) and on the usefulness score (coefficient = $0.136$ and $0.156$ respectively), all with a statistical significance at the level of $p < 0.05$. The existence of a \textit{decoy} also seems to render a positive impact on duration (coefficient = $1.916$ and $1.913$ respectively), however, the result on both datasets is not statistically significant. 

More precisely, the regression results can be interpreted as follows: Given the document rank, type of task, and individual characteristics, when a \textit{decoy} is present, in comparison to when it is absent (1)~the likelihood of being clicked on the THUIR2016 and THU-KDD datasets would respectively increase by $36.3\%$ ($p<0.05$) and $21.7\%$ ($p<0.05$); (2)~browsing time duration would rise by $1.92$s  and $1.91$s respectively on the THUIR2016 and THU-KDD datasets; (3)~usefulness score would escalate by $0.136$ ($p<0.01$) and $0.156$ ($p<0.01$) respectively on the THUIR2016 and THU-KDD datasets.

\input{tables/reg}

%% file: fig_data_process.tex
\begin{figure*}[t]
    \centering
    \includegraphics[width=.85\linewidth]{
    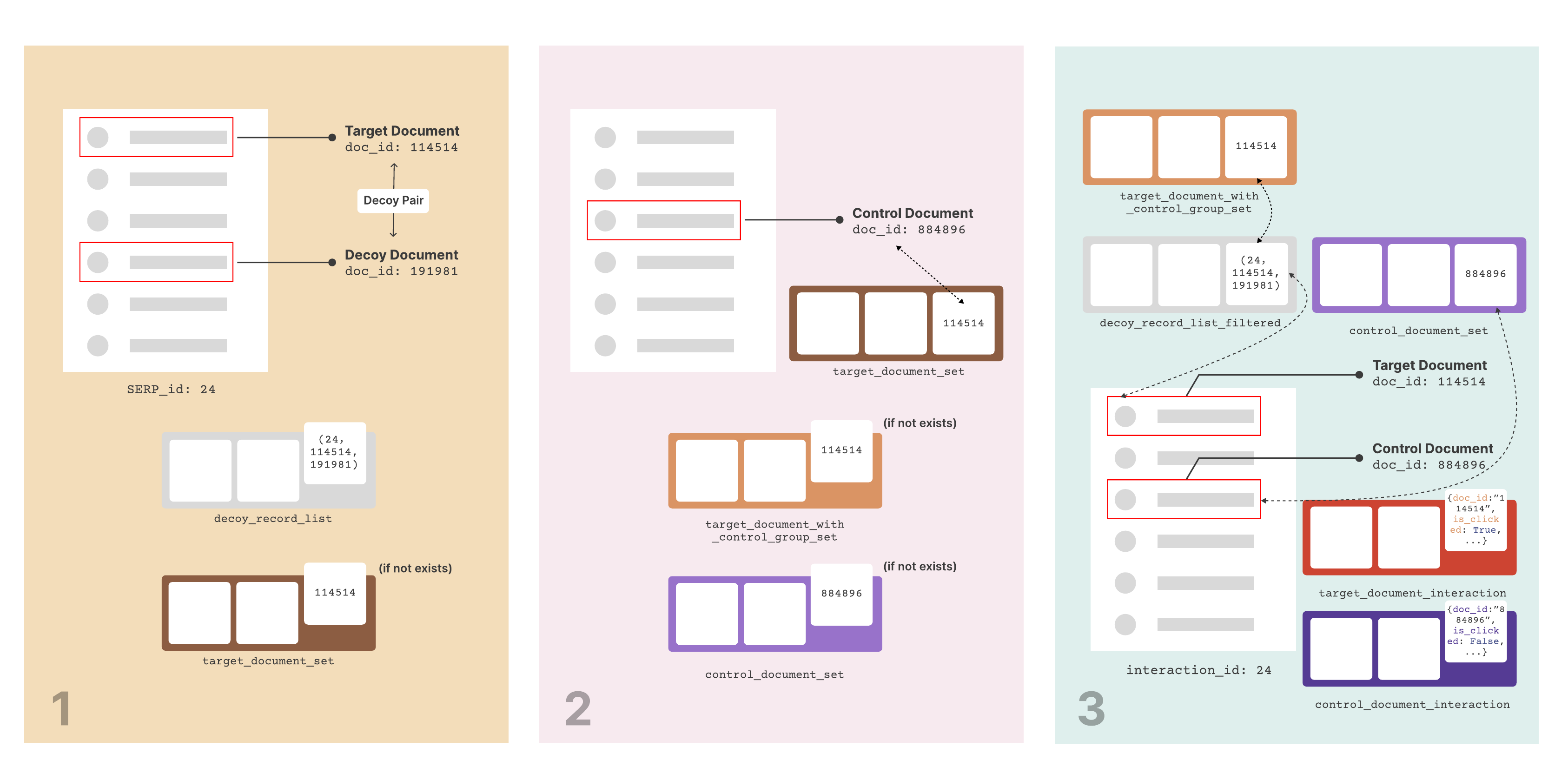
    }
    \caption{\label{fig:data_process} Data processing flow in the first experiment.}
\end{figure*}

%% file: fig_feat.tex
\begin{figure*}[htbp]
\centering
\begin{subfigure}[t]{0.33\textwidth}
    \centering
    \includegraphics[width=5.5cm]{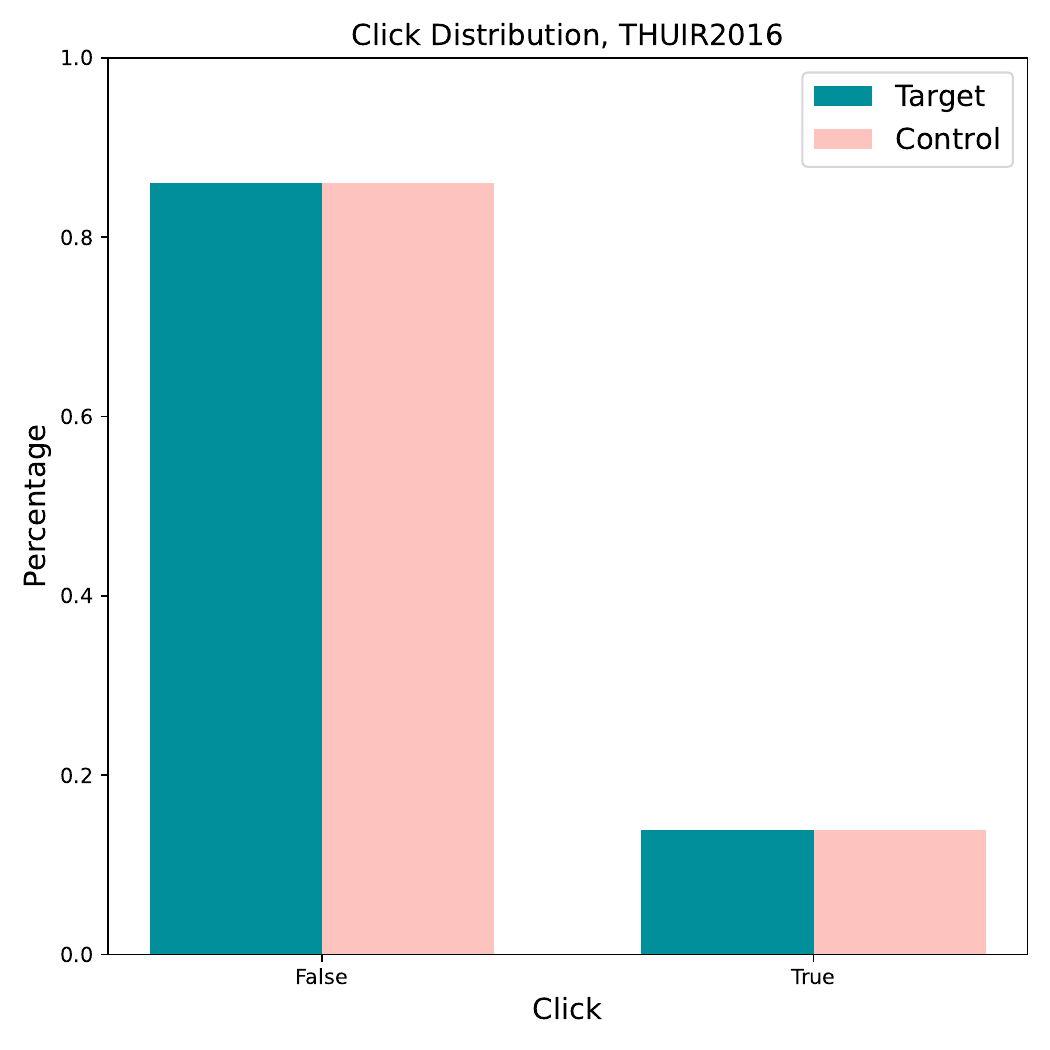}
    \caption*{}
\end{subfigure}
\begin{subfigure}[t]{0.33\textwidth}
    \centering
    \includegraphics[width=5.5cm]{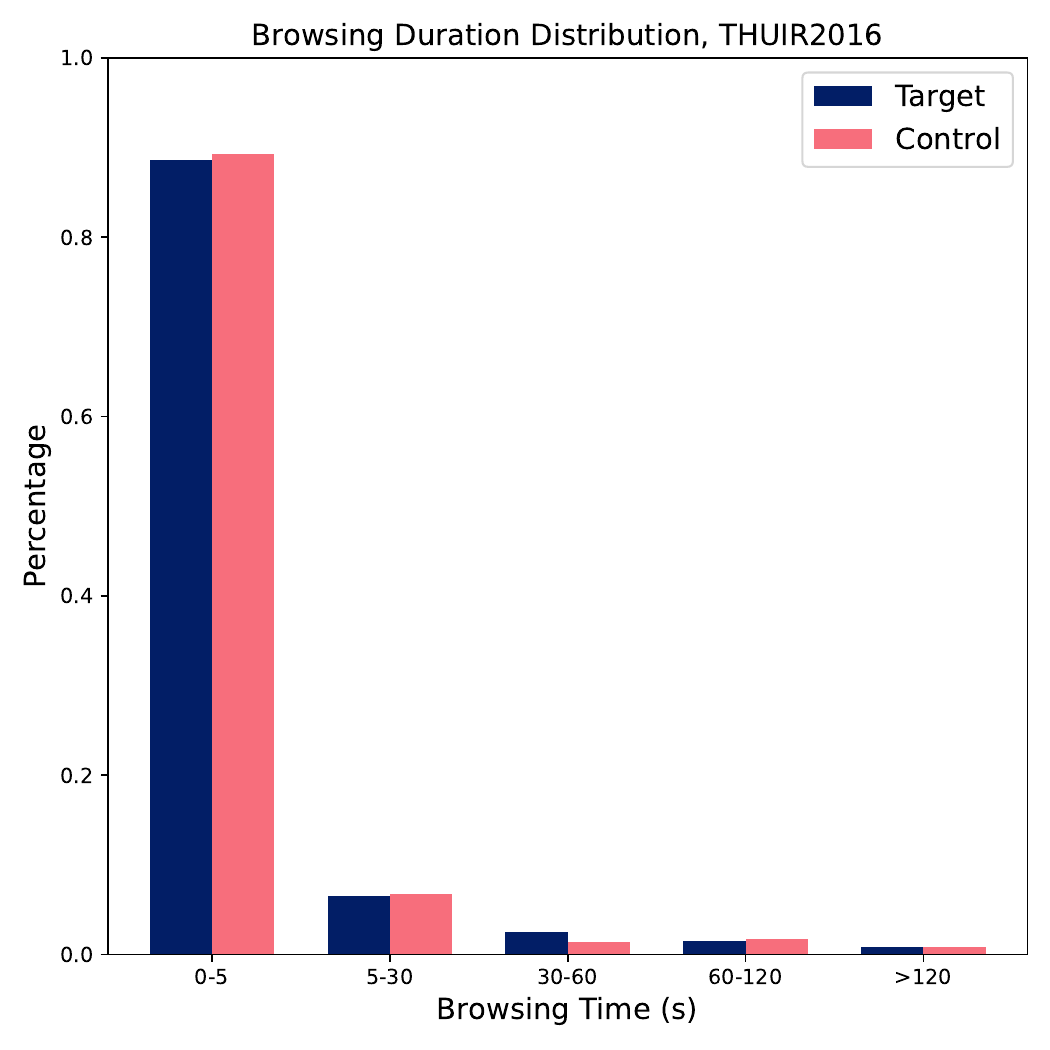}
    \caption*{}
\end{subfigure}
\begin{subfigure}[t]{0.33\textwidth}
    \centering
    \includegraphics[width=5.5cm]{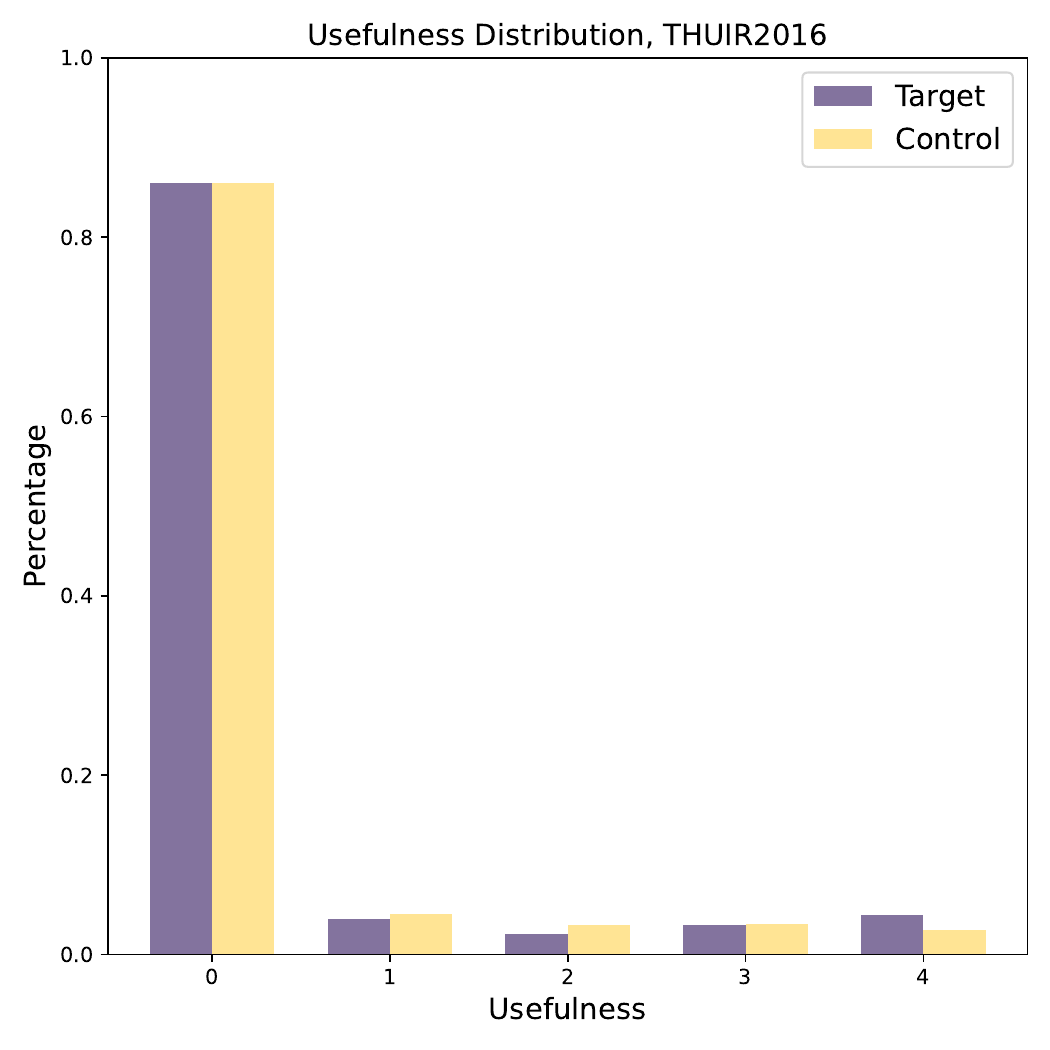}
    \caption*{}
\end{subfigure}
\begin{subfigure}[t]{0.33\textwidth}
    \centering
    \includegraphics[width=5.5cm]{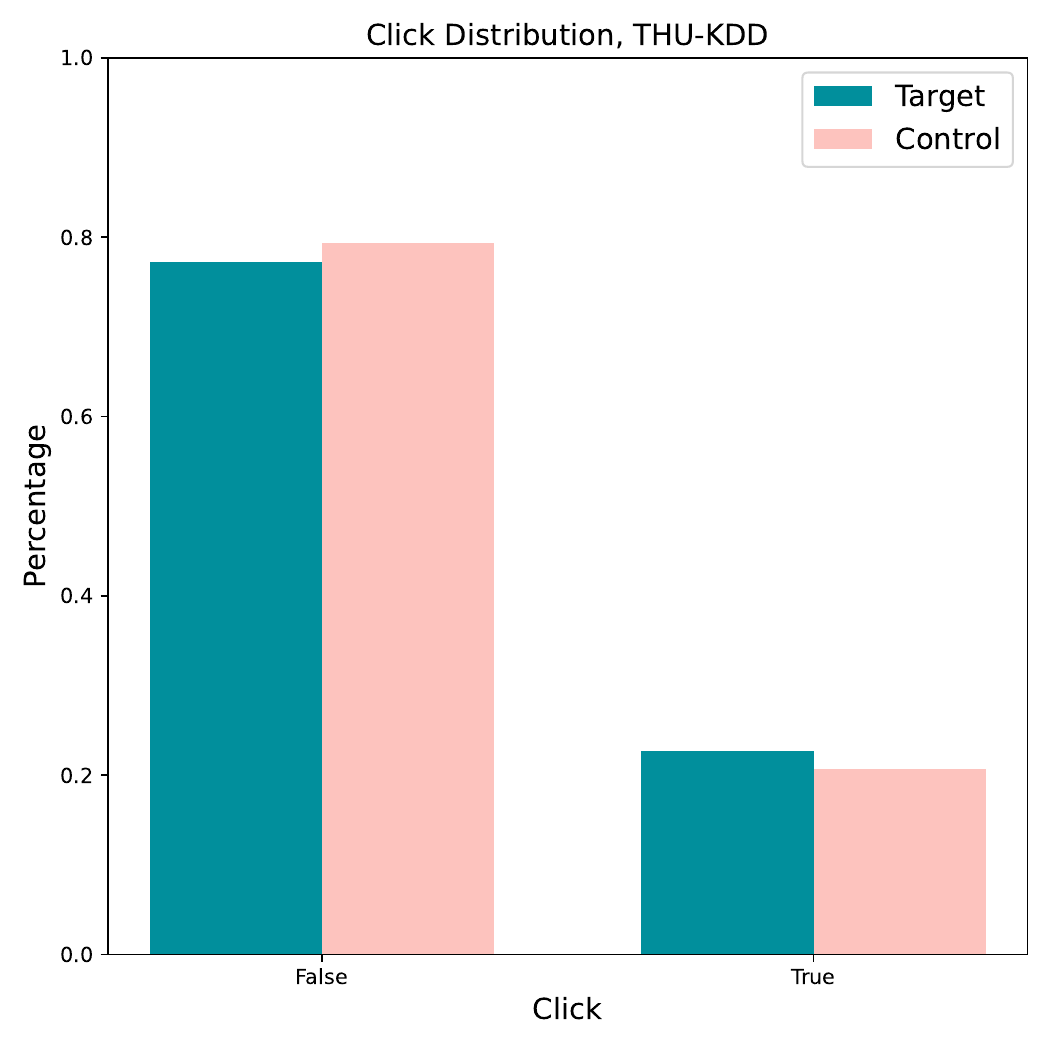}
    \caption*{}
\end{subfigure}
\begin{subfigure}[t]{0.33\textwidth}
    \centering
    \includegraphics[width=5.5cm]{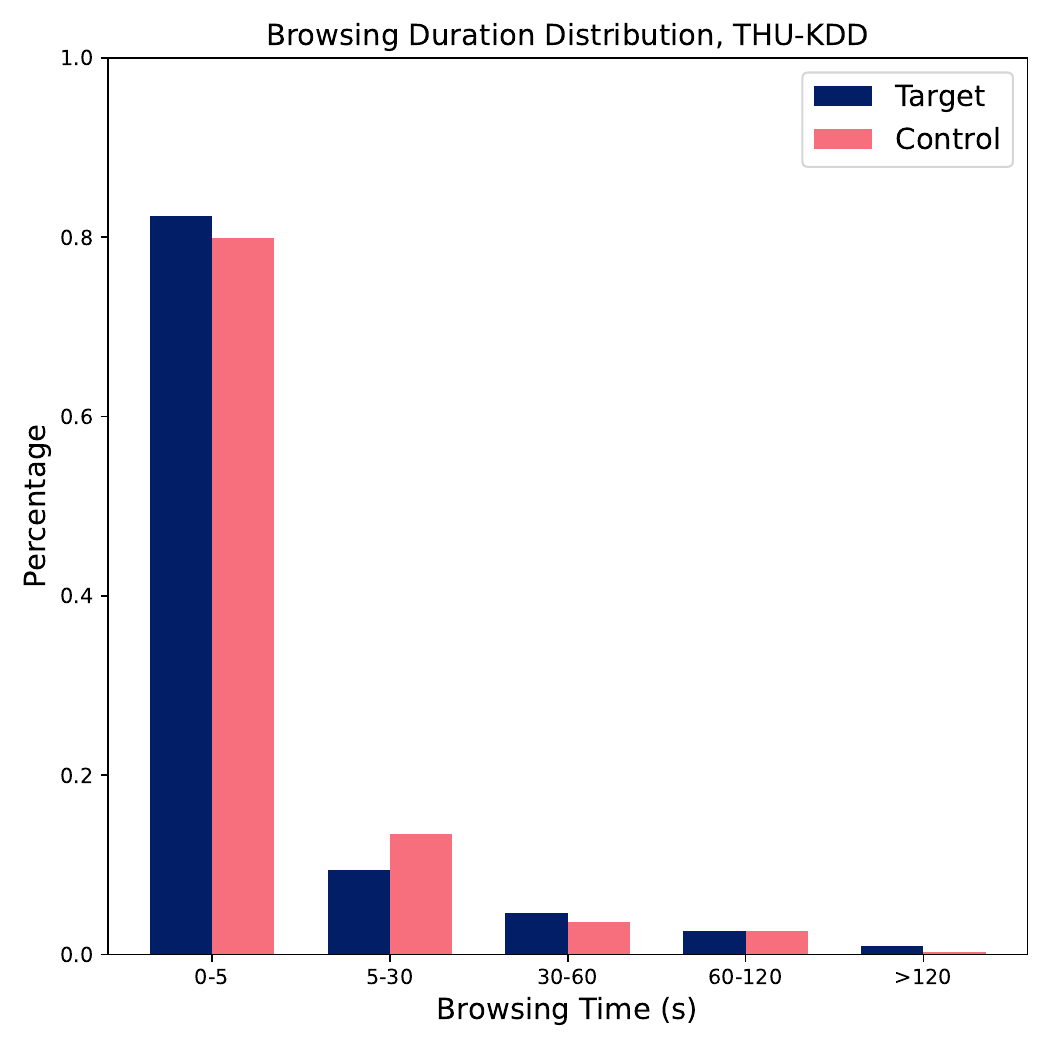}
    \caption*{}
\end{subfigure}
\begin{subfigure}[t]{0.33\textwidth}
    \centering
    \includegraphics[width=5.5cm]{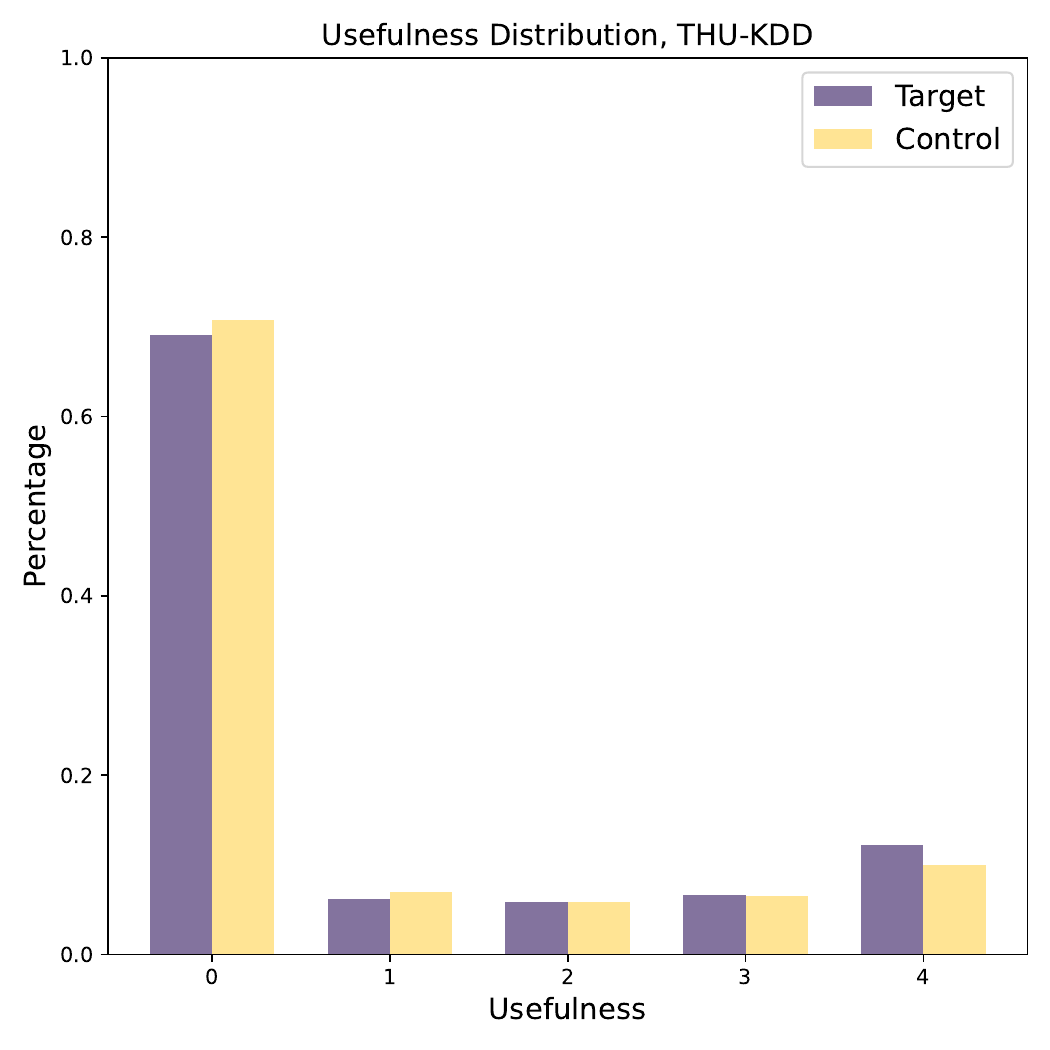}
    \caption*{}
\end{subfigure}
\caption{The distribution of click probability, browsing time and usefulness score on THUIR2016 dataset (top) and THU-KDD (bottom) dataset respectively.}
\Description{}
\label{fig:feat}
\end{figure*}

%% file: fig_dist.tex
\begin{figure*}[htbp]
\centering
\begin{subfigure}[t]{0.49\textwidth}
    \centering
    \includegraphics[width=8.2cm]{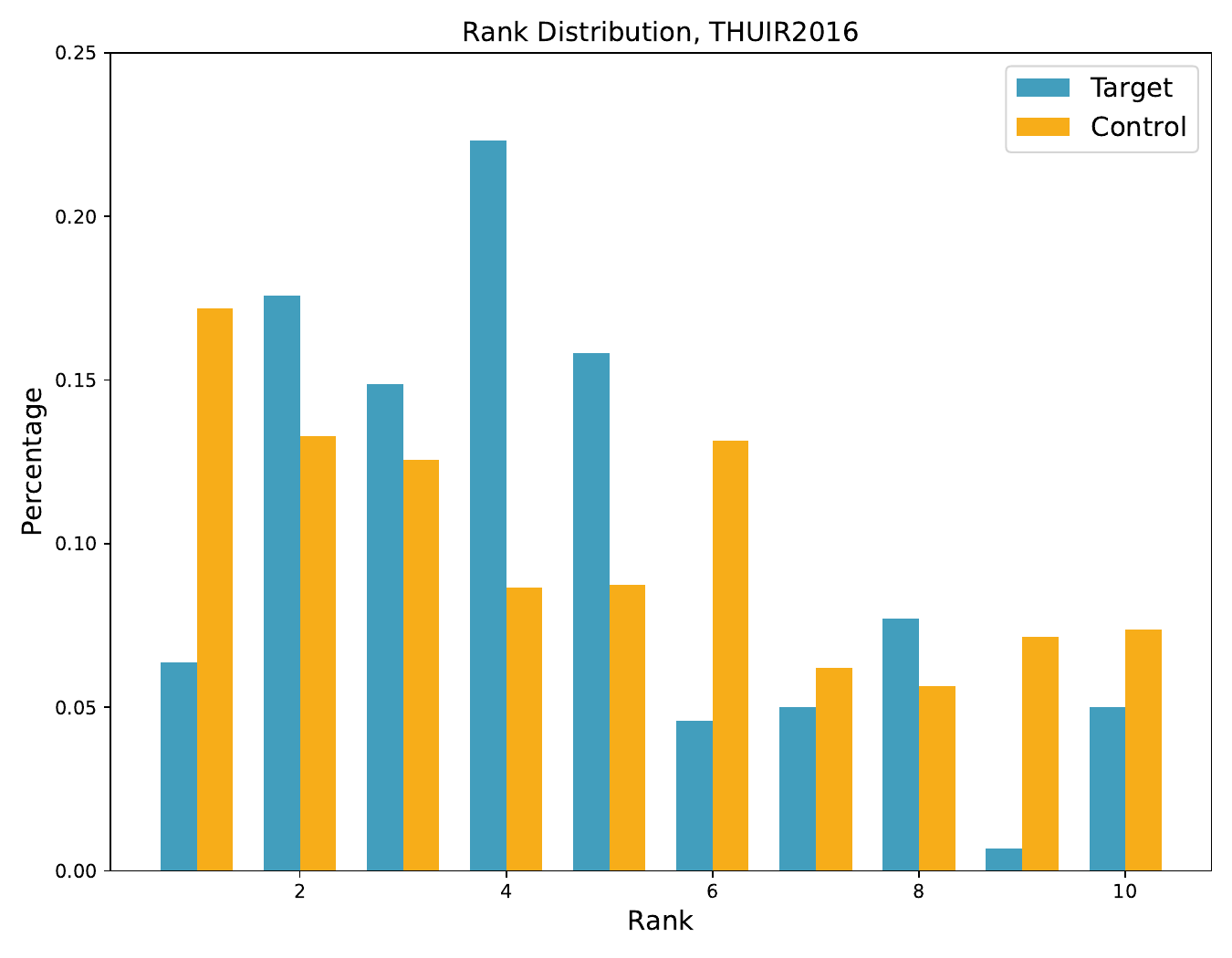}
    \caption*{}
\end{subfigure}
\begin{subfigure}[t]{0.49\textwidth}
    \centering
    \includegraphics[width=8.2cm]{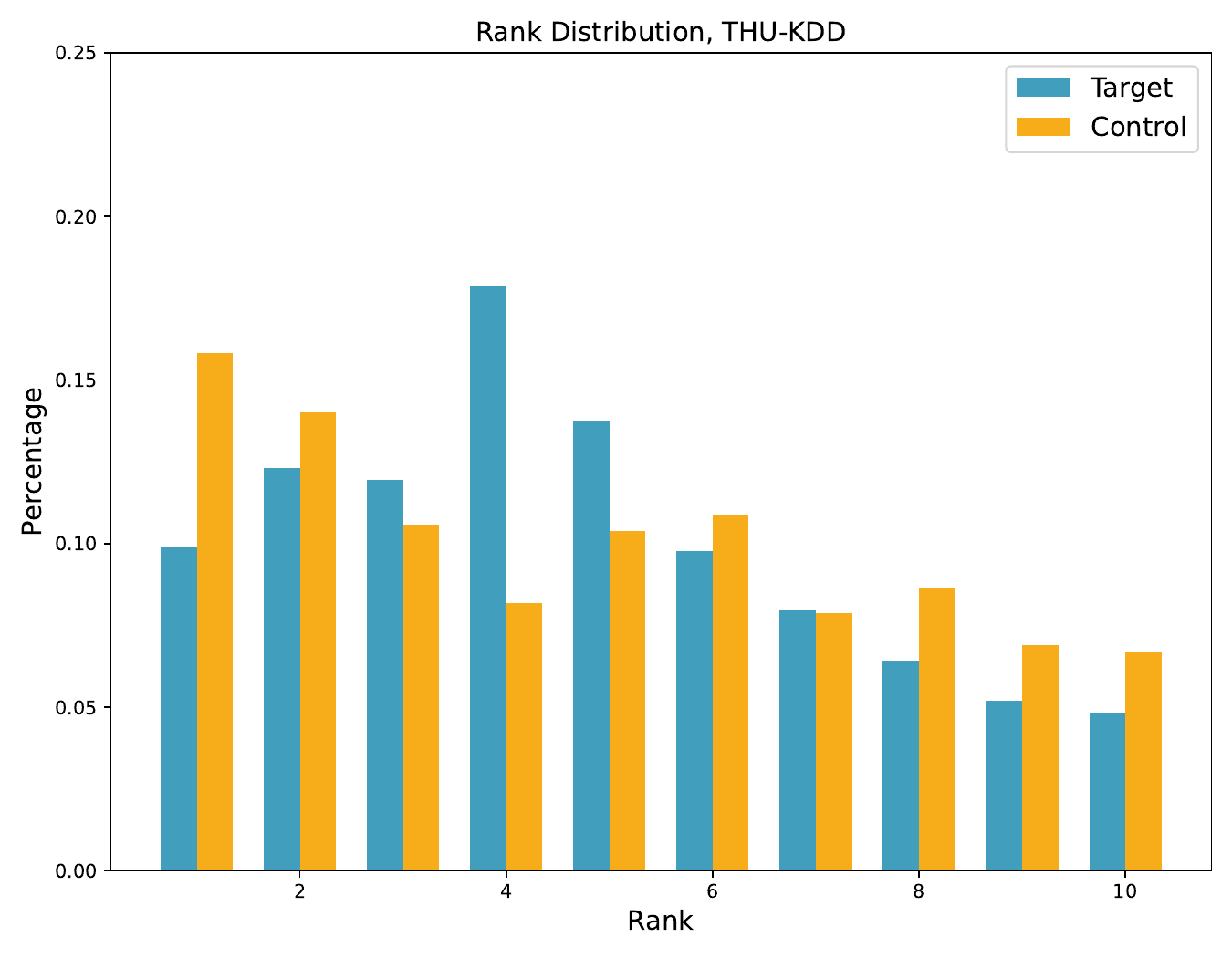}
    \caption*{}
\end{subfigure}
\caption{The distribution of rank on THUIR2016 (left) dataset and THU-KDD (right) dataset respectively.}
\Description{}
\label{fig:dist}
\end{figure*}

%% file: tables/reg.tex
\begin{table}[htbp]
\centering
\begin{tabular}{ccc}
\hline
  & THUIR2016 & THU-KDD \\
 Dependent Variable & Coefficient & Coefficient \\
 \hline
 \texttt{is\_clicked} & 0.363* & 0.217* \\
 \texttt{duration}  & 1.916 & 1.913 \\
 \texttt{usefulness}  & 0.136 ** & 0.156 ** \\
\hline
\end{tabular}
\caption{The regression coefficient of the independent variable \texttt{has\_decoy} with the dependent variables \texttt{is\_clicked}, \texttt{duration}, and \texttt{usefulness}. * and ** respectively indicate that the coefficients are significant at levels of \( p < 0.05 \), \( p < 0.01 \).}
\label{table:reg}
\end{table}

%% file: conclusion.tex
\section{Conclusion}

In this study, we seek to comprehend how the \textit{decoy
effect} at the document level impacts users’ interaction behaviors
on SERPs, such as clicks, dwell time, and usefulness perceptions.
We conducted descriptive data analysis and regression analysis on two publicly available user behavior datasets. The experimental results indicate that,  the likehood of a document being clicked could be improved and its usefulness score could be elevated, should there be a decoy associated with the document. From the experimental results, we observe that, given the document rank, type of task and individual characteristics, when a decoy is present, in comparison to when it is absent, there is a significant increment in the likehood of a document being clicked and its perceived usefulness.

As far as we know, we are the first to addresses how the decoy effect influences user interactions on Search Engine Result Pages. Our work extends the endeavors of the IR community in exploring how cognitive biases impact user behaviors in document examining and relevance judgment, providing evidence from the perspective of the decoy effect.

However, our study is merely in a preliminary stage, with numerous aspects awaiting further exploration. For instance:~(1)~Does the impact of the decoy vary in size under different themes, topics, or search tasks?~(2)~Within the same search session, does the impact of decoy results vary across different cognitive states and information seeking intentions (e.g. exploring an unfamiliar domain, seeking for a known item, evaluating retrieved information)~\cite{liu2020identifying}?~(2)~The potential impact of the decoy effect on the benefit of users and information providers, as well as possible social and ethical issues it may bring about, such as fairness. Addressing these questions will necessitate subsequent experiments to provide more evidence.